\documentstyle[preprint,aps]{revtex}

\begin{document}
\draft
\preprint{}
\title{New Algorithm for Mixmaster Dynamics \thanks{
e-mail: berger@oakland.edu, garfinkl@oakland.edu}}
\author{Beverly K. Berger, David Garfinkle and Eugene Strasser}
\address{Department of Physics, Oakland University, Rochetster, MI 48309}
\date{\today}
\maketitle
\bigskip
\begin{abstract}
We present a new numerical algorithm for evolving the Mixmaster
spacetimes.  By using symplectic integration
techniques to take advantage of the exact Taub solution for the scattering between
asymptotic Kasner regimes, we evolve these spacetimes with higher accuracy using
much larger  time steps than previously possible. The longer Mixmaster
evolution thus allowed enables detailed comparison with the Belinskii,
Khalatnikov, Lifshitz (BKL) approximate Mixmaster dynamics. In particular, we
show that errors between the BKL prediction and the measured parameters early
in the simulation can be eliminated by relaxing the BKL assumptions to yield an
improved map. The improved map has different predictions for vacuum Bianchi
Type IX and magnetic Bianchi Type VI$_0$ Mixmaster models which are clearly
matched in the simulation.
\end{abstract}
\pacs{98.80.Dr, 04.20.J}


Mixmaster dynamics (MD), discovered independently by Belinskii, Khalatnikov,
and Lifshitz (BKL)
\cite{bkl} and Misner \cite{misner}, describes a system in which pure gravity
exhibits chaos (or at least a strong sensitivity to initial conditions).
Although first discovered in vacuum spatially homogeneous cosmologies of
Bianchi Type IX, MD also occurs in vacuum cosmologies of Bianchi Type VIII
\cite{halpern} and magnetic Bianchi Type VI$_0$ \cite{LKW}. (Other possible
arenas for MD are given by Jantzen \cite{jantzen}.) It is clear that the most
general {\it homogeneous} cosmology exhibits MD
and has been conjectured that the same is true in
the inhomogeneous case  \cite{bkl}. Much recent work \cite{hobill} has focused on
the question of whether or not MD is chaotic in any invariant sense since
computed Lyapunov exponents can be either zero or positive depending on the
choice of time variable \cite{rugh,berger1,ferraz}. The issue has recently been
resolved in favor of chaos by Cornish and Levin
\cite{cornish} who have provided a prescription to define the discrete outcomes
required to exhibit the fractal basins of attraction characteristic of chaos. A
parallel issue remains, however. BKL (as revised by Chernoff and Barrow
\cite{bc} and extended by Berger \cite{berger2}) derived an approximate MD as a
sequence of Kasner models with a map from one Kasner epoch to the next. So far,
the properties of the map have always been consistent with those of the full
solution to Einstein's equations obtained numerically
\cite{rugh,berger1,berger2}. However, one would wish to have more precise
criteria for the BKL map's validity and perhaps to measure departures from it.
In addition, one must untangle the loss of information due to the chaotic nature
of the dynamics from the accumulation of errors due to finite numerical
precision.

Here, we describe a new numerical algorithm for MD which represents
improvement by at least two orders of magnitude over standard ODE solvers in
the speed with which a Mixmaster model may be evolved toward the singularity
(without any loss of accuracy). We take advantage of the fact that MD as a
sequence of Kasners is equivalent to MD as a sequence of bounces (transitions
between Kasners) and that evolution from one Kasner to the next is the exactly
solvable Taub cosmology \cite{taub}. We apply this new method to the (diagonal)
Bianchi IX and magnetic Bianchi VI$_0$ examples of MD in order to compare an
improved BKL map to the numerical results and to display clearly the role of
numerical precision in this comparison.

The models we shall use to illustrate our method are described by the metric
$$
d {s^2} = - \, {e^{2 (\alpha + \zeta + \gamma )}} \, d {t^2} \; + \; {e^{2 \alpha }} \, {{\left ( {\sigma _1} \right ) }^2} \; + \; {e^{2 \zeta }} \, {{\left ( {\sigma _2} \right ) }^2} \; + \; {e^{2 \gamma }} \, {{\left ( {\sigma _3} \right ) }^2} \; \; \; .
\eqno(1)
$$
Here $\alpha , \, \zeta $ and $\gamma $ are functions of $t $, and 
$ {\sigma _1} , \, {\sigma _2} $ and $\sigma _3$ are time independent,
orthogonal forms, invariant under the group of spatial symmetries.  
The dynamics of these spacetimes is
therefore just the determination of the logarithmic scale factors (LSFs)
$\alpha , \, \zeta $ and $\gamma $ as functions of $t $.  The Einstein equations
for the models may be obtained by variation of the Hamiltonian
$H = {H_k} \; + \; {H_p}$
(and the constraint $H=0$) where
$$
2 \, {H_k} =  3 \, \left ( {p_\alpha ^2} \, + \, {p_\zeta ^2} \, + \,{p_\gamma ^2} \right ) \; - \; 6 \, \left ( {p_\alpha } {p_\zeta } \, + \,  {p_\alpha } {p_\gamma } \, + \,  {p_\zeta } {p_\gamma } \right ) \; \; \; .
\eqno(2)
$$
The potential term $H_p$ is a function of the LSFs and depends on the type of
homogeneous cosmology being treated.  For vacuum Bianchi type IX or magnetic
Bianchi Type VI$_0$, we have
$$
H_p = {c^2 \, e^{2 b \alpha}} \, + \,  {e^{4 \zeta}} \, + \,  {e^{4
\gamma}}
\; -
\; 2 \, \left ( a \, {e^{2 (\alpha + \zeta )}} \, + \, a \, {e^{2 (\alpha +
\gamma )}}
\, + \, d \, {e^{2 (\zeta + \gamma )}} \right )   \; \; \; .
\eqno(3)
$$
Here $a = 1$, $b=2$, $c=1$ and
$d = 1$ for vacuum Bianchi Type IX, while  $a = 0$, $b=1$, $c = {\sqrt \xi} $
and $ d = -1$ for magnetic Bianchi Type VI$_0$,   
and $\xi $ is a constant that depends on the strength of the magnetic field.
Note that the magnetic field in Type VI$_0$ provides the potential wall that is
furnished by curvature in Type IX. When the logarithmic scale factors are large
and negative, the potential term
$H_p$ is negligible so that the dynamical system is then essentially a free
particle: $\alpha , \, \zeta $ and $\gamma $ are linear functions of $t$.
This is a Kasner epoch.  The bounces between epochs occur in those periods when the potential is not negligible.  Since the potential consists of terms that
are exponentials of the LSFs, each of the bounces occurs
in a brief time period (in terms of $t$).

To evolve the Mixmaster spacetimes numerically we use the method of symplectic
integration \cite{fleck,suzuki}.  This method begins with a dynamical system
with Hamiltonian $H={H_1} + {H_2}$ where the equations obtained by variation of
$H_1$ and
$H_2$ are separately exactly solvable. For a
Hamiltonian $H$ let $U(H,\Delta t)$ be the operator that evolves the system for
a time $\Delta t$.  Consider the operator
$$
{U_2} (H, \Delta t) \equiv U({H_1},\Delta t/2) \,  U({H_2},\Delta t) \, U({H_1},\Delta t/2) \; \; \; .
\eqno(4)
$$
then to second order in $\Delta t$ the operator ${U_2} (H, \Delta t)$ agrees
with $U(H,\Delta t)$.  Using the operator ${U_2} (H, \Delta t)$ one can,
by iteration,
construct evolution operators that agree with $U(H,\Delta t)$ to any
desired order.  Explicitly the $2n+2$ order approximation to $U(H,\Delta t)$ is
$$
{U_{2n+2}} (H, \Delta t) \equiv {U_{2n}}(H,{s_n} \Delta t) \,  {U_{2n}}(H,(1 - 2 {s_n})\Delta t) \, {U_{2n}}(H,{s_n} \Delta t) 
\eqno(5)
$$
where ${s_n} = 1/(2 - {2^{1/(2n+1)}})$.  
Note that though evolution using $U_{2n}$ is formally $2n$ order accurate, the
error may be much smaller than one would expect.  Recall that if ${H_2} = 0$
then $U_2$ (and therefore $U_{2n}$) is exactly equal to $U$.  Similarly if the 
evolution takes place in a region of phase space where $H_2$ is extremely small
then $U_{2n}$ is an extremely good approximation to $U$, a much better
approximation than, {\it e.g.} $2n$ order Runge-Kutta.  Thus in such regions of
phase space one can take very large time steps without introducing large
inaccuracy.

To apply the symplectic integration method to the homogeneous cosmologies it
is necessary to divide their Hamiltonian into two pieces, each of
which is exactly solvable.  It turns out that $H_k$ and $H_p$ are each exactly
solvable since $H_k$ is a free particle Hamiltonian that yields the Kasner
solution while $H_p$ contains no momenta and thus yields trivial equations of
motion.  However, splitting the Hamiltonian in this way still necessitates the
use of extremely small time steps at the bounces.  To take full advantage of the
symplectic method we would like to split the Hamiltonian in such a way that
$H_2$ is very small in the region of phase space in which the evolution is
taking place.  Note that the largest contribution to the potential term is
always of the form $ {c^2} {e^{2 b {\tilde
\alpha }}} $ where $b$ and
$c$ are the constants defined in (3) for $\tilde \alpha$ corresponding to a
magnetic wall and $c=1$, $b=2$ for $\tilde \alpha$ corresponding to a curvature
wall. Call the other two LSFs $\tilde \zeta$ and $\tilde \gamma$. 
Denote by
$ {{\tilde p}_\alpha} ,
\, {{\tilde p}_\zeta} $ and ${\tilde p}_\gamma$ the momenta corresponding to 
$ {\tilde \alpha } , \, {\tilde \zeta}$ and $\tilde \gamma$ respectively.
We split the Hamiltonian into 
${H_1}={H_k} + {c^2} {e^{2 b {\tilde \alpha }}}$ and
${H_2}={H_p} - {c^2} {e^{2 b {\tilde \alpha }}}$. But $H_1$ is the Hamiltonian
for the exactly solvable Taub model while $H_2$ still contains no momenta. At
each time step, the code identifies the largest LSF and implements the
appropriate split.

In our notation, the Taub solution is the following:
Since $H_1$ is independent of $\tilde \zeta $ and $\tilde \gamma $ 
it follows that 
${\tilde p}_\zeta $ and ${\tilde p}_\gamma $ are constants.  
The remaining variables evolve as follows:  
Define $q, k$ and $r$ by $q \equiv - \, {\dot {\tilde \alpha}} (t)$,  
$$
k \equiv {{\left [ {q^2} \, + \, 6 \, {c^2} \,  {e^{2 b \tilde \alpha (t)}}
\right ] }^{1/2}} \; \; \; ,
\eqno(6)
$$
$$
r \equiv \ln \left ( \cosh k b \Delta t \; + \; {q \over k} \; \sinh k b \Delta t \right ) \; \; \; .
\eqno(7)
$$
Then we have
$$
\tilde \alpha ( t + \Delta t) = \tilde \alpha (t) \; - \; {r \over b}  \; \; \; ,
\eqno(8)
$$
$$
{{\tilde p}_\alpha } ( t + \Delta t) = {{\tilde p}_\alpha } (t) \; - \; {{2
{c^2}}
\over k}
\; {e^{2 b \tilde \alpha (t) - r}} \, \sinh k b \Delta t \; \; \; ,
\eqno(9)
$$
$$
\tilde \zeta ( t + \Delta t) = \tilde \zeta (t) \; + \; {r \over b}  \; - \; 6
{{\tilde p}_\gamma }
\,
\Delta t \; \; \; ,
\eqno(10)
$$
$$
\tilde \gamma ( t + \Delta t) = \tilde \gamma (t) \; + \; {r \over b}  \; - \; 6
{{\tilde p}_\zeta }
\, \Delta t \; \; \; .
\eqno(11)
$$
Note that these formulas give exact solutions of the dynamics of $H_1$.  There
is no approximation and no assumption that $\Delta t$ is ``small.''  The
dynamics of $H_2$ are trivial.  Since $H_2$ is independent of 
${{\tilde p}_\alpha } , \,
{{\tilde p}_\zeta }$ and ${\tilde p}_\gamma$ it follows that 
${\tilde \alpha} , \, {\tilde \zeta}$ and $\tilde \gamma$
are constants.  The evolution of ${\tilde p}_\alpha $ is then given by
$$
{{\tilde p}_\alpha } ( t + \Delta t) = {{\tilde p}_\alpha } (t) \; - \; {{\partial {H_2}} \over {\partial {\tilde \alpha }}} \; \Delta t 
\eqno(12)
$$
and correspondingly for the other momenta.  

Our computer code implements the sixth order symplectic integration algorithm
with this split of the Hamiltonian.  The time step is increased if evolution
with $\Delta t$ and $\Delta t / 2$ yields the same solution to a desired
accuracy. For comparison we also evolved the same Mixmaster spacetimes with a
fourth order, adaptive stepsize Runge-Kutta code and a sixth order symplectic
algorithm with
$H=H_k+H_p$. A comparison of the results of the three codes for type IX is shown 
in Fig.~1 for the region around a typical bounce.  Note that the symplectic
integration code achieves accurate evolution in approximately one hundred times
fewer steps than the Runge-Kutta code. The key here is that any adaptive step
size algorithm in any ODE solver can accurately reproduce the Kasner solution.
However, in standard methods, $\Delta t$ must be drastically decreased when a
potential term becomes non-negligible. If care is not taken, a too large time
step will cause overshoot of the potential wall leading to numerical disaster.
In the new algorithm, the largest term in (3) is identified and the appropriate
Taub solution used to effortlessly propagate the trajectory through the bounce.
There is no need to decrease the time step. One can easily reach $|\Omega| =
{1 \over 3} |\alpha + \zeta + \gamma| \approx 10^{35}$ in only 50,000 time
steps corresponding to more than 150 epochs. Previous simulations have
reported results for fewer than 30 epochs with $| \Omega | < 10^8$ while
requiring significantly more computer time \cite{berger1}. If the constraint is
not enforced at all, none of the algorithms discussed here can evolve more than
about 15 epochs with the initial data used here. To maintain accuracy, it is
necessary to enforce the Hamiltonian constraint at least every few time steps
depending on the precision required. Recently, Gundlach and Pullin
\cite{gundlach} have argued that the constraints must be enforced in generic
numerical relativity codes. Our new algorithm allows simulations to be run for a
sufficiently large number of epochs that the need to enforce the constraint here
is also seen.

Each Kasner epoch can be specified by the values of four parameters: 
($u,v,{p_\Omega},\kappa$). As shown by several authors\cite{berger3,rugh} these
parameters can be defined throughout the evolution of a Mixmaster spacetime,
though they are (approximately) constant only within a Kasner epoch.  Following
reference\cite{berger3} we define
${p_\Omega } \equiv {p_\alpha} + {p_\zeta} + {p_\gamma} \; , \;$and the Kasner
indices $  {p_1}= - {\dot
\alpha}/(3 {p_\Omega}) \; , \; {p_2}= - {\dot
\zeta}/(3 {p_\Omega})$ and $ {p_3}= - {\dot \gamma}/(3 {p_\Omega}) $. Now
reorder the $p_i$ so that ${p_1} \le {p_2} \le {p_3} $.  Let ${\bar \alpha } ,
\, {\bar \zeta }$ and $\bar \gamma$ be respectively the LSFs corresponding to
${p_1} , \, {p_2} $ and $p_3$.  Define $u, \, v$ and $\kappa$ by
$$
u \equiv - \, 1 \; - \; {{p_3} \over {p_1}} \; \; \; ,
\eqno(13)
$$
$$
v \equiv {{p_2} \over {p_3}} \; {{\left ( {p_1} {\bar \gamma} \, - \; 
{p_3} {\bar \alpha} \right ) } \over {\left ( {p_1} {\bar \zeta} \, - \; 
{p_2} {\bar \alpha} \right ) }} \; + \; 1 \; \; \; ,
\eqno(14)
$$
$$
\kappa = {p_3} \; \left ( {{\bar \zeta} \over {p_2}} \; - \; {{\bar \alpha} \over {p_1}} \right ) \; \; \; .
\eqno(15)
$$
BKL made the following approximation to 
MD: (i) between bounces the spacetime is exactly 
Kasner, (ii) the bounce occurs instantaneously and (iii) at a bounce one
of the LSFs vanishes.  This approximation yields a rule (the BKL map) that
gives the values of the parameters in the $n+1$st epoch in terms of their
values in the $n$th epoch.  It is well-known \cite{hobill} that these assumptions
are less valid early in the simulation and improve as the singularity is
approached. For $p_\Omega$, the BKL rule is 
$$
{p_{\Omega , n+1}} = {p_{\Omega , n}} \; \left ( {{{u_n ^2} \, - \, {u_n} \, + \, 1} \over {{u_n ^2} \, + \, {u_n} \, + \, 1}} \right ) \; \; \; .
\eqno(16)
$$
For the other parameters, the BKL map depends on whether $u_n$ is greater than
$2$.  For ${u_n} \ge 2$, the BKL map is ${u_{n+1}} = {u_n} - 1 , \; {v_{n+1}} =
{v_n} + 1 $ and ${\kappa _{n+1}} ={\kappa _n}$.  For ${u_n} \le 2 $  (referred to
as an era change), the rule is $ {u_{n+1}} = 1/({u_n}-1) , \;
{v_{n+1}}=1+(1/{v_n})$ and ${\kappa _{n+1}} = {v_n} {\kappa _n} /({u_n}-1)$.
However, the known exact dynamics of $H_1$ provides a better approximation to
MD than does this BKL approximate dynamics---i.e.~the sequence
of bounces is a better description than the sequence of Kasners.  It therefore
yields a better map of the parameters ($u,v,{p_\Omega},\kappa$) from one epoch
to the next.  This  ``improved BKL map'' actually agrees with the BKL map for
$u$ and $p_\Omega$ but adds corrections to $v$ and $\kappa$.  Let the quantity
$w$ be given by
$$
w = {2 \over b} \; \ln \left ( {c \over {\sqrt 6}} \; {{{u^2} + u + 1} 
\over { u \, {p_\Omega}}} \right ) \; \; \; 
\eqno(17)
$$
where $b$ and $c$ are defined as in (3) for the vacuum Type IX and magnetic
VI$_0$ cases. Within any era the improved BKL map for
$v$ and $\kappa$ is
$$
{v_{n+1}} = {v_n} \; + \; 1 \; + \; {{{u_n} + 1 - {v_n}} \over {1 + ({\kappa _n} / {w_n} )}} \; \; \; ,
\eqno(18)
$$
$$
{\kappa _{n+1}} = {\kappa _n} \, + \, {w_n} \; \; \; ,
\eqno(19)
$$
while for an era change the improved BKL map gives 
$$
{v_{n+1}} = 1 \; + \; {{1 \, + \, ({w_n} / {\kappa _n} )} \over {{v_n} \, + \,
({u_n} + 1) \, ({w_n}/{\kappa _n})}} \; \; \; ,
\eqno(20)
$$
$$
{\kappa _{n+1}} = {{{v_n} {\kappa _n} \, + \, ({u_n} + 1) {w_n}} \over {{u_n} 
- 1 }} \; \; \; .
\eqno(21)
$$
Note that in the limit of vanishing $w_n/\kappa_n$ the improved BKL map agrees
with the BKL map.  As the singularity is approached, $|\kappa_n| \to \infty$ so
that the BKL terms overwhelm the corrections. In Fig.~2, 20 epochs of a typical
magnetic Bianchi VI$_0$ trajectory are displayed in the anisotropy plane. The
axes are
\cite{moser} $\beta_{\pm} / |\Omega|$ where $\beta_+ = {1 \over 6}(\zeta +
\gamma -2 \alpha)$, $\beta_- = {1 \over {2 \sqrt{3}}} (\zeta + \gamma)$ so that
the exponential potential terms have fixed locations. We have chosen the
vertical wall to be due to the magnetic field ($\xi e^{2 \alpha}$) with the
other two walls ($e^{4 \zeta}$, $e^{4 \gamma}$) due to curvature. The epoch
numbers are identical to those in Fig.~3 which shows the fractional difference
between measured and predicted (on the basis of the previous measured value)
values of $v$ for the BKL and improved maps for bounces off magnetic (VI) and
curvature (IX) walls. It is clearly seen that extremely accurate agreement is
obtained alternately for the magnetic and curvature improved maps as the
trajectory correspondingly bounces off the respective walls. A new era with
bounces between two curvature walls is also seen in epochs 12-14. Fig.~4 shows a
later stage in the evolution when all three maps converge as expected. The
results for the parameter $\kappa$ are identical.

Finally, we consider tracking an initial value of $u$ through
almost 200 epochs with the Hamiltonian constraint enforced at every time
step. The sequence is begun at a point where, during an era, the ODE
solution follows the BKL map for $u$ to the desired precision. Unlike the
predictions previously described, loss of information and numerical error are
allowed to accumulate. {\it Mathematica}
\cite{mathematica} is used to obtain the sequence of BKL $u$-values to 60
significant figures (SFs) while our ODE solver is run in quadruple precision
(32 SFs). Integer parts for $u$ in the latter different from those predicted by
the former arise in our simulation at epoch 161, precisely where a 32 SF {\it
Mathematica} sequence completely loses precision due to era changes. In fact,
examination of the
$u$ sequence shows that epoch 161 begins the 32nd era, as expected. (Similarly,
the double precision simulation deviates at epoch 47, the start of the 16th
era.) Given initial data specified to $N$ SFs, the numerical simulation
follows the true Mixmaster trajectory
through $\approx N$ eras, until the initial information is lost. However, as
can be deduced from Fig.~4, the numerical solution remains close to some true
Mixmaster trajectory for any sequence of $N$ eras in the simulation. We emphasize
here that these effects are not visible early in the simulation. Our new
algorithm allows easy study of Mixmaster models for more than one hundred epochs
to permit careful examination of issues relating to the validity of the
(improved) BKL approximation.

\section*{Acknowledgments}
This work was supported in part by 
National Science Foundation Grants PHY9507313 (BKB) and PHY9408439 (DG) to
Oakland University (OU). DG was also supported in part by a Cottrell College
Science Award of Research Corporation to OU. BKB would like to
thank the Institute for Geophysics and Planetary Physics at Lawrence Livermore
National Laboratory and the Department of Astronomy at the University of
Michigan for hospitality. Part of this work was performed by ES in partial
fulfillment of the requirements for the M.S.~in Physics degree at OU.

%
%

\begin{figure}
\caption{Comparison of algorithms for MD. A typical Mixmaster
bounce is shown in the anisotropy plane. Crosses indicates every 10th point on a
4th order Runge-Kutta evaluation of the trajectory, while circles indicate every
10th point for a 6th order standard symplectic evaluation. The filled squares
indicate {\it every} point using the new algorithm. The inset shows the details
closest to the bounce.}
\end{figure}

\begin{figure}
\caption{The first 20 epochs of a typical magnetic Bianchi VI$_0$ 
trajectory in the rescaled anisotropy plane where the potential is shown as a
fixed equilateral triangle. Some epoch numbers are shown. The vertical potential
wall is produced by the magnetic field while the others are due to curvature.}
\end{figure}

\begin{figure}
\caption{Fractional difference between measured and predicted values of $v$ vs.
epoch number using the BKL prediction (crosses), the curvature wall prediction
(squares), and the magnetic wall prediction (circles). Triangles mark the first
epoch of an era.}
\end{figure}

\begin{figure}
\caption{Convergence of the BKL and improved maps as the evolution shown in
Figures 2 and 3 is continued toward the singularity. The symbols are the same
as those in Fig.~3.}
\end{figure}

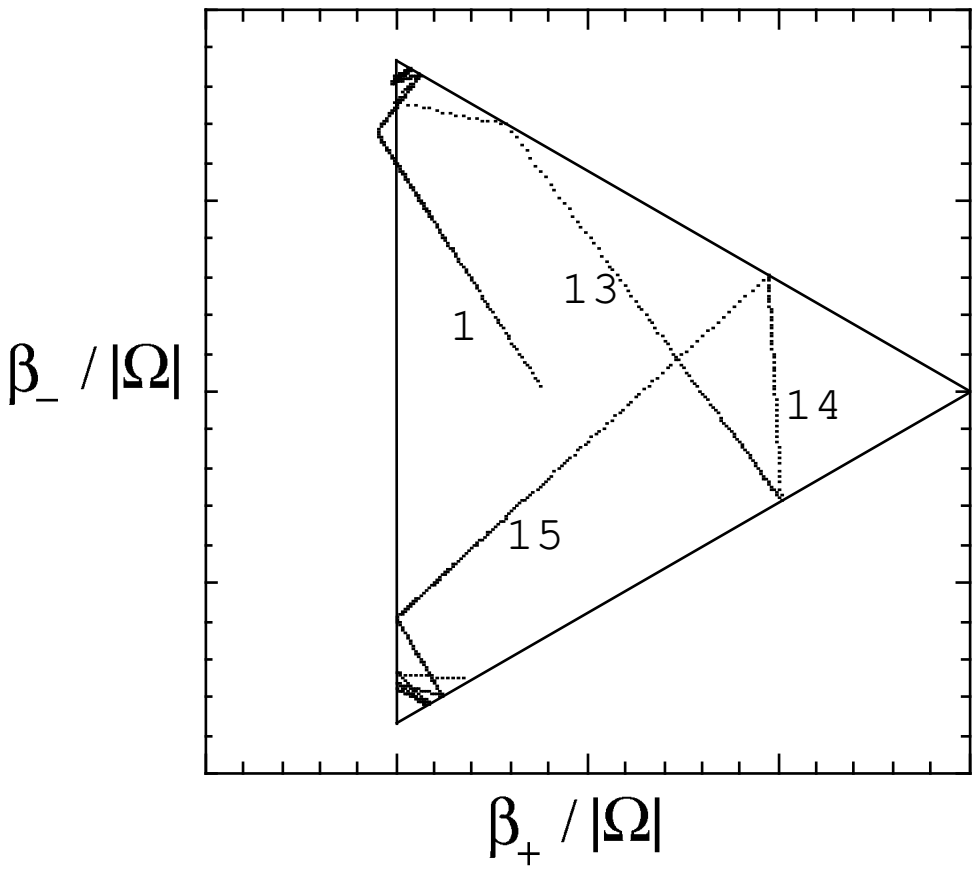
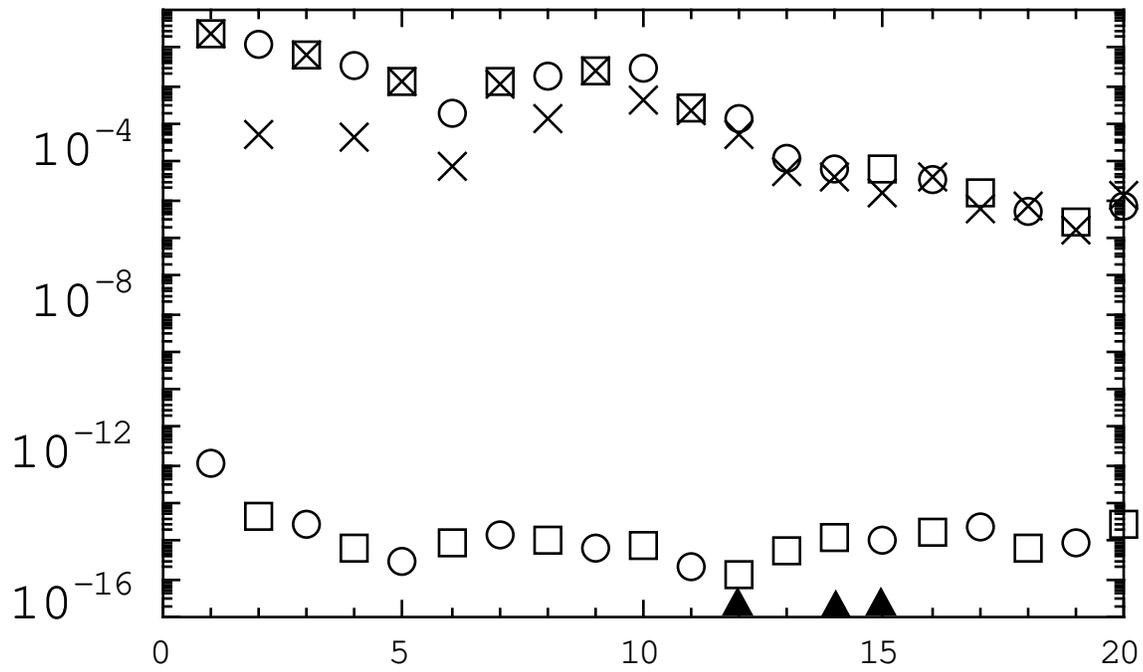
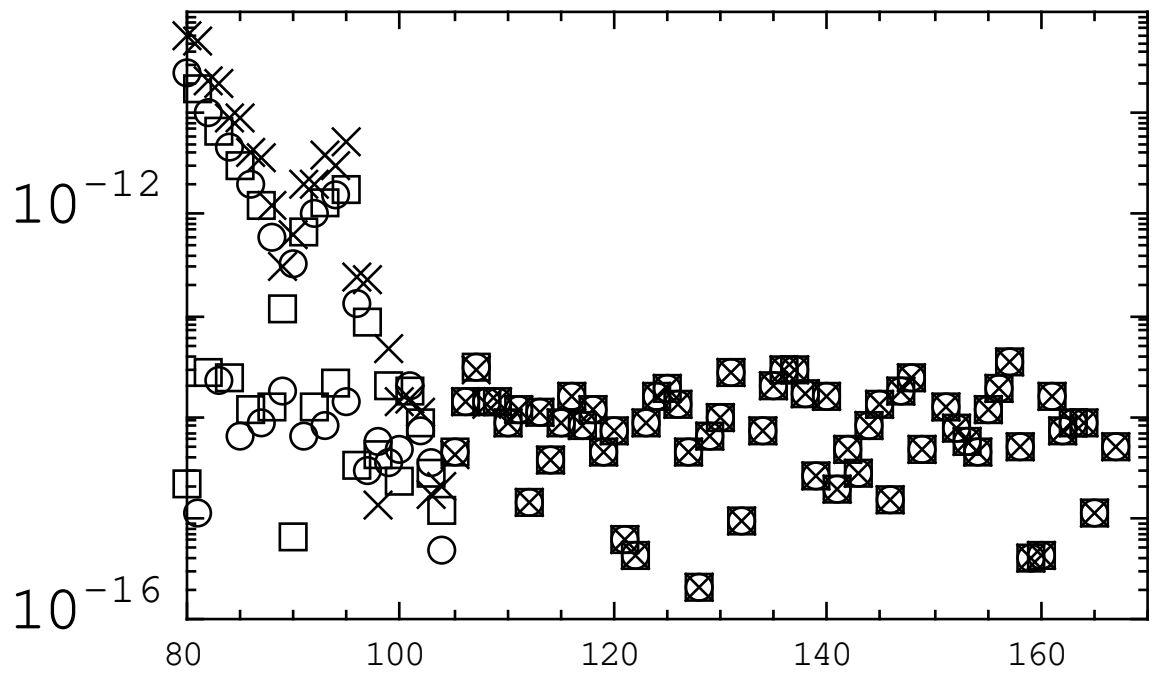
\end{document}